\documentclass[12pt,english]{extarticle}
\usepackage[T1]{fontenc}
\usepackage[latin9]{inputenc}
\usepackage{geometry}
\usepackage{color}
\usepackage{babel}
\usepackage{amsmath}
\usepackage{cancel}
\usepackage{graphicx}
\usepackage[unicode=true]
 {hyperref}

\makeatletter

\providecommand{\tabularnewline}{\\}

\newcommand{\lyxaddress}[1]{
	\par {\raggedright #1
	\vspace{1.4em}
	\noindent\par}
}

\date{}
\usepackage{hyperref, setspace, lineno}
\usepackage{lmodern}
\usepackage{soul} 

\hypersetup{
    colorlinks=true,      
    linkcolor=red,          
    citecolor=blue,        
    filecolor=magenta,      
    urlcolor=blue           
}

\@ifundefined{showcaptionsetup}{}{%
 \PassOptionsToPackage{caption=false}{subfig}}
\usepackage{subfig}
\makeatother

\begin{document}
\title{Control of Dendrites in Rechargeable Batteries using Smart Charging}
\author{Asghar Aryanfar$^{\dagger}$$^{\ddagger}$\thanks{Corresponding author; email: \protect\href{mailto:aryanfar@caltech.edu}{aryanfar@caltech.edu}}
, Yara Ghamlouche$^{\dagger}$, William A. Goddard III$^{\S}$}
\maketitle

\lyxaddress{\begin{center}
\emph{$\dagger$ American University of Beirut, Riad El-Solh, Lebanon
1107 2020 }\\
\emph{$\ddagger$ Bahçe\c{s}ehir University, 4 Ç\i ra\u{g}an Cad,
Be\c{s}ikta\c{s}, Istanbul, Turkey 34353}\\
\emph{$\S$ California Institute of Technology, 1200 E California
Blvd, Pasadena, CA 91125}
\par\end{center}}
\begin{abstract}
In this paper we develop a feed-back control framework for the real-time
minimization of microstructures grown within the rechargeable battery.
Due to quickening nature of the branched evolution, we identify the
critical ramified peaks in the early stages and based on the state
we compute the relaxation time for the concentration in those branching
fingers. The control parameter is a function of the maximum curvature
(i.e. minimum radius) of the branched microstructure, where the higher
rate dendritic evolution would lead to the more critical state to
be controlled. The charging time is minimized for generating the most
packed microstructures and obtained results correlate closely with
those of considerably higher charging time periods. The developed
framework could be utilized as a smart charging protocol for the safe
and sustainable operation the rechargeable batteries, where the branching
of the microstructures could be correlated to the sudden variation
in the current/voltage.
\end{abstract}

\section{Introduction}

The modern era of wireless revolution and portable electronics demands
the utilization of reliable intermittent renewables and long-lasting
electrical energy storage facilities \cite{Rugolo12,Dunn11}. As well,
the growing demand for portable computational power as well as the
introduction of electric vehicles demand novel and reliable high capacity
energy storage devices. Despite such impressive growth of the need
in the daily lifestyle, the underlying science remains to be developed.
Rechargeable batteries, which retrieve/store energy from/within the
chemical bonds, have proven the be the most reliable and cleanest
resource of electrical energy for the efficient management of the
power. \cite{Marom11,Goodenough13} Metallic electrodes such as lithium
$Li$ \cite{Cheng17}, sodium $Na$ \cite{Slater13}, magnesium $Mg$
\cite{Davidson18}, and zinc $Zn$ \cite{Pei14} are arguably highly
attractive candidates for use in high-energy and high-power density
rechargeable batteries. Lithium $Li$ possess the lowest mass density
($\rho_{Li}=0.53~g.cm^{-3}$) and the highest electropositivity ($E^{0}=-3.04V$
vs SHE\footnote{$SHE$: Standard Hydrogen Electrode, taken conventionally as the reference
($E_{H^{2}}^{0}=0$)}) which provides the highest gravimetric energy density and likely
the highest voltage output, making it suitable for high-power applications
such as electric vehicles \cite{Li19Lithium,Xu14,Zhamu12}. Sodium
has a lower cost and is more earth abundant and is operational for
large-scale stationary energy storage applications \cite{Ellis12}.
Magnesium $Mg$ possess a high specific capacity and reactivity \cite{Shterenberg14}
whereas Zinc $Zn$ is earth-abundant, has low cost and high storage
capacity \cite{Li14Recent}.

During the charging, the fast-pace formation of microstructures with
relatively low surface energy from Brownian dynamics, leads to the
branched evolution with high surface to volume ratio \cite{Xu14}.
The quickening tree-like morphologies could occupy a large volume,
possibly reach the counter-electrode and short the cell. Additionally,
they can also dissolve from their thinner necks during subsequent
discharge period and form detached dead crystals, leading to thermal
instability and capacity decay \cite{Xu04,Tewari20}. The quickening
tree-like morphologies could occupy a large volume, possibly reach
the counter-electrode and short the cell. Additionally, they can also
dissolve from their thinner necks during subsequent discharge period.
Such a formation-dissolution cycle is particularly prominent for the
metal electrodes due to lack of intercalation\footnote{Intercalation: diffusion into inner layer as the housing for the charge,
as opposed to depositing in the surface.}, where the depositions in the surface is the only dominant formation
mechanism versus the diffusion into the inner layers as the housing
\cite{Dresselhaus81,Li14Review} The growing amorphous crystals can
pierce into the polymer electrolyte and short the cell afterwards,
given their higher porosity, they could have mechanical properties
comparable to the bulk form \cite{Aryanfar17Bulk}.

Previous studies have investigated various factors on dendritic formation
such as current density \cite{Ren15}, electrode surface roughness
\cite{Nielsen15,Natisiavas16}, impurities \cite{Steiger14}, solvent
and electrolyte chemical composition \cite{Schweikert13,Younesi15},
electrolyte concentration \cite{Jeong08}, utilization of powder electrodes
\cite{Seong08} and adhesive polymers\cite{Stone12}, temperature
\cite{Aryanfar15Annealing,Aryanfar15Thermal}, guiding scaffolds \cite{Yao19,Qian19},
capillary pressure \cite{Deng19}, cathode morphology \cite{Abboud19}
and mechanics \cite{Klinsmann19,Xu17,Wang19,Liu19Piezo}. Some of
conventional characterization techniques used include NMR \cite{Bhattacharyya10}
and MRI. \cite{Chandrashekar12} Recent studies also have shown the
necessity of stability of solid electrolyte interphase (i.e. SEI)
layer for controlling the nucleation and growth of the branched medium
\cite{Li19Energy,Kasmaee16} as well as pulse charging \cite{Li01,Chandrashekar16,Aryanfar18}.

Earlier model of dendrites had focused on the electric field and space
charge as the main responsible mechanism \cite{Chazalviel90} while
the later models focused on ionic concentration causing the diffusion
limited aggregation (DLA). \cite{Monroe03,Witten83,Zhang19,Fleury97}
Both mechanisms are part of the electrochemical potential \cite{Bard80,Tewari19},
indicating that each could be dominant depending on the localizations
of the electric potential or ionic concentration within the medium.
Recent studies have explored both factors and their interplay, particularly
in continuum scale and coarser time intervals, matching the scale
of the experimental time and space \cite{Aryanfar14Dynamics}. Other
simplified frameworks include phase field modeling \cite{Mu19,Ely14,Gogswell15}
and analytical developments \cite{Akolkar13}.

During charge period the ions accumulate at the dendrites tips (unfavorable)
due to high electric field in convex geometry and at the same time
tend to diffuse away to other less concentrated regions due to diffusion
(favorable). Such dynamics typically occurs within the double layer
(or stern layer \cite{Bazant11}) which is relatively small and comparable
to the Debye length. In high charge rates, the ionic concentration
is depleted on the reaction sites and could tend to zero~\cite{Fleury97};
Nonetheless, our continuum-level study extends to larger scale, beyond
the double layer region \cite{Aryanfar19Finite}

Dendrites instigation is rooted in the non-uniformity of electrode
surface morphology at the atomic scale combined with Brownian ionic
motion during electrodeposition. Any asperity in the surface provides
a sharp electric field that attracts the upcoming ions as a deposition
sink. Indeed the closeness of a convex surface to the counter electrode,
as the source of ionic release, is another contributing factor. In
fact, the same mechanism is responsible for the further semi-exponential
growth of dendrites in any scale. During each pulse period the ions
accumulate at the dendrites tips (unfavorable) due to high electric
field in convex geometry and during each subsequent rest period the
ions tend to diffuse away to other less concentrated regions (favorable).
The relaxation of ionic concentration during the idle period provides
a useful mechanism to achieve uniform deposition and growth during
the subsequent pulse interval. Such dynamics typically occurs within
the double layer (or stern layer \cite{Bazant11}) which is relatively
small and comparable to the Debye length. In high charge rates, the
ionic concentration is depleted and concentration on the depletion
reaches zero~\cite{Fleury97}; Nonetheless, our continuum-level study
extends to larger scale, beyond the double layer region \cite{Aryanfar14Dynamics}. 

Pulse method has been qualitatively proved as a powerful approach
for the prevention of dendrites \cite{Li01}, which has previously
been utilized for uniform electroplating \cite{Chandrashekar08}.
In the preceding publication we have experimentally found that the
optimum rest period correlates well with the relaxation time of the
double layer for the blocking electrodes which is interpreted as the
\emph{RC time} of the electrochemical system \cite{Bazant04}. We
have explained qualitatively how relatively longer pulse periods with
identical duty cycles $\mathbf{D}$ will lead to longer and more quickening
growing dendrites. We developed coarse grained computationally affordable
algorithm that allowed us reach to the experimental time scale (\emph{$ms$}).
We have developed theoretical limit the optimal minimization of the
dendrites \cite{Aryanfar18} and we have obtained the pulse charging
parameters for individual curved peaks based on their curvature \cite{Aryanfar19Finite}. 

In this paper, we elaborate on the real-time controlling of the pulse
charging parameters for the minimization of microstructures grown
in the scales extending to the cell domain. We analyze the oscillatory
behavior and the transition from initial to steady-state growth regimes. 

\section{Methodology}

\begin{figure}
\noindent\begin{minipage}[t]{1\columnwidth}%
\begin{minipage}[t]{0.48\textwidth}%
\begin{center}
\includegraphics[height=0.24\textheight]{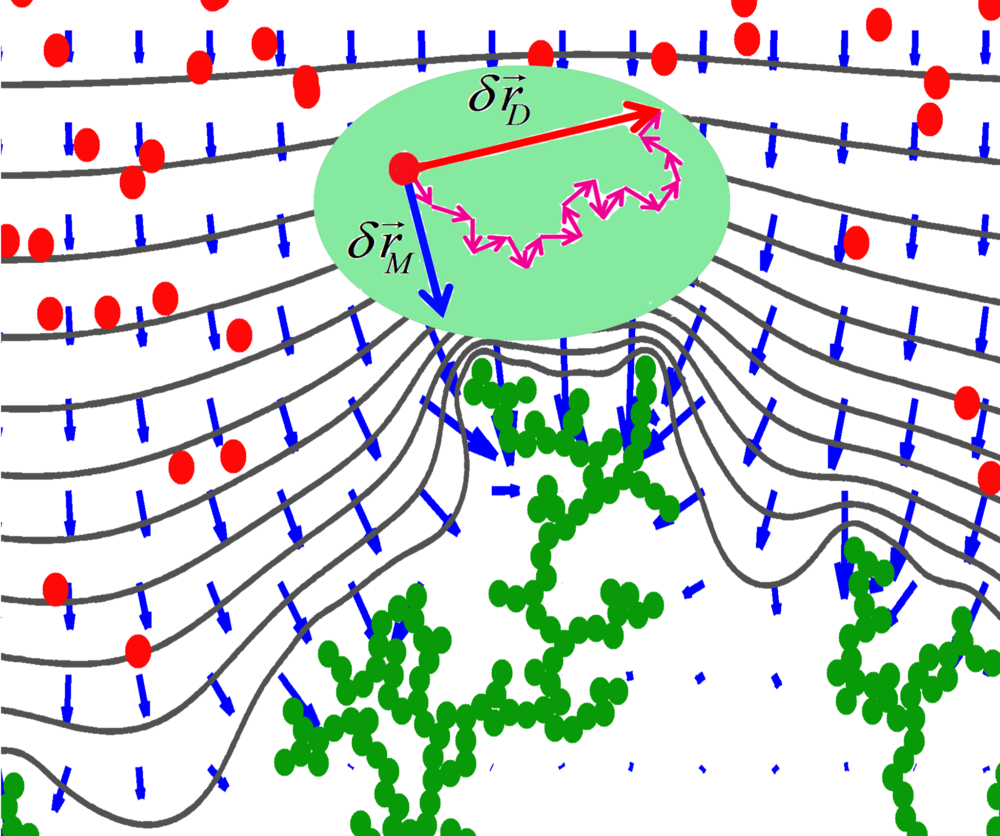}
\par\end{center}
\caption{The transport elements in the coarse scale of time and space.\label{fig:Displacements}}
\end{minipage}\hfill{}%
\begin{minipage}[t]{0.48\textwidth}%
\begin{center}
\includegraphics[height=0.24\textheight]{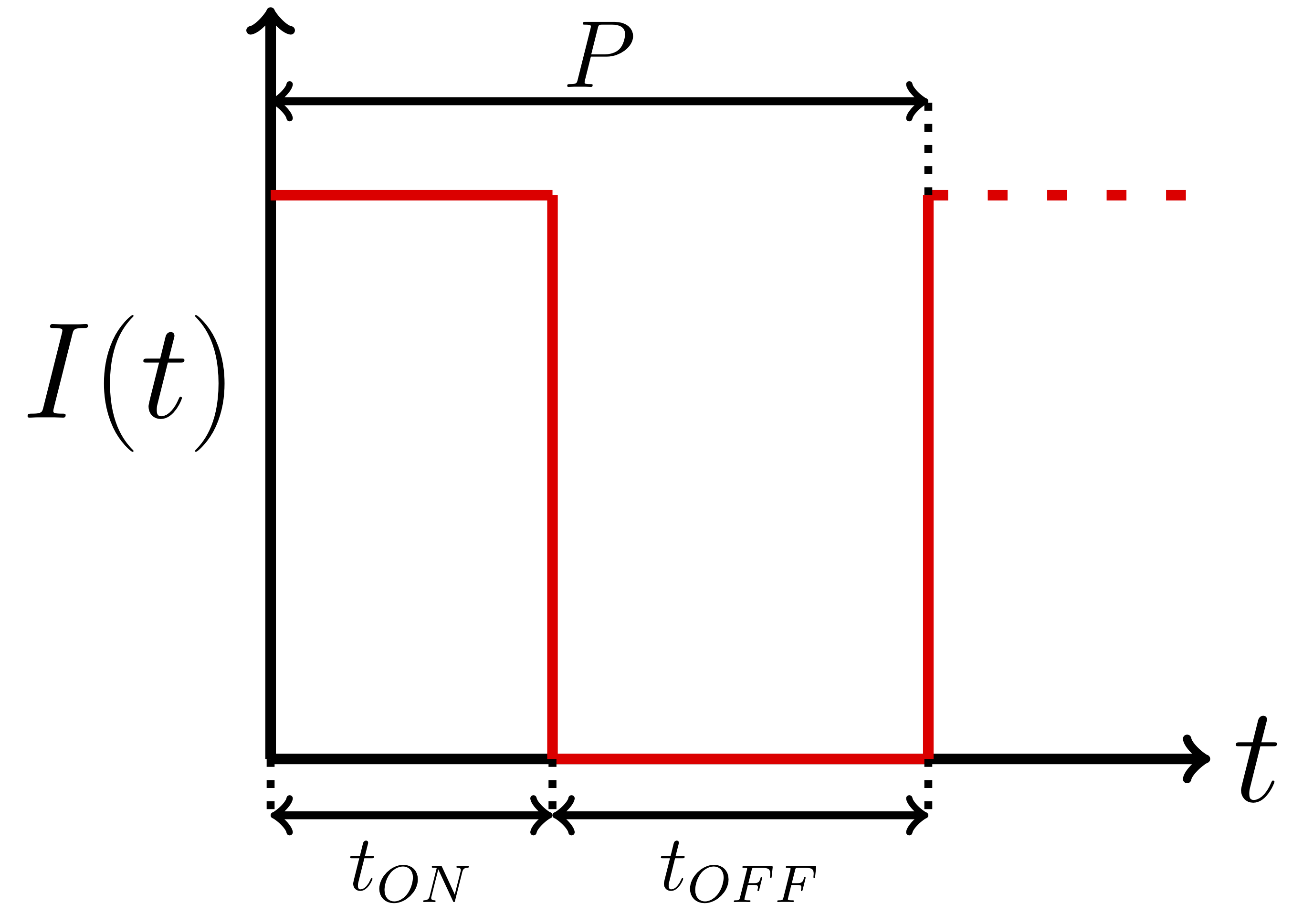}
\par\end{center}
\caption{Square pulse wave.\label{fig:PulseCurve}}
\end{minipage}%
\end{minipage}
\end{figure}

The electrochemical flux is generated either from the gradients of
concentration ($\nabla C$) or electric potential ($\nabla V$). In
the ionic scale, the regions of higher concentration tend to collide
and repel more and, given enough time, diffuse to lower concentration
zones, following Brownian motion. In the continuum (i.e. coarse) scale,
such inter-collisions could be added-up and be represented by the
diffusion length $\delta\vec{\textbf{r}}_{D}$ as: \cite{Aryanfar14Dynamics}
\footnote{\textcolor{black}{The diffusion coefficient $D^{+}$ is generally
concentration dependent \cite{Chandrashekar16}, due to electro-neutrality
within the considerable space in the domain and we assume it is constant
in the range considered.}}

\begin{equation}
\delta\vec{\textbf{r}}_{D}=\sqrt{2D^{+}\delta t}\text{ }\hat{\textbf{g}}\label{eq:DiffDis}
\end{equation}

where $\vec{\textbf{r}}_{D}$ is diffusion displacement of individual
ion, $D^{+}$ is the ionic diffusion coefficient in the electrolyte,
$\delta t$ is the coarse time interval \footnote{$\delta t=\sum_{i=1}^{n}\delta t_{i}$ where $\delta t_{k}$ is the
inter-collision time, typically in the range of $fs$.}, and $\hat{\textbf{g}}$ is a normalized vector in random direction,
representing the Brownian dynamics. The diffusion length represents
the average progress of a diffusive wave in a given time, obtained
directly from the diffusion relationship \cite{Philibert06}. 

On the other hand, ions tend to acquire drift velocity in the electrolyte
medium when exposed to electric field and during the given time $\delta t$
their progress $\delta\vec{\textbf{r}}_{M}$ is given as:

\begin{equation}
\delta\vec{\textbf{r}}_{M}=\mu^{+}\vec{\textbf{E}}\delta t\label{eq:MigDis}
\end{equation}
where $\mu^{+}$ is the mobility of cations in electrolyte, $\vec{\textbf{E}}$
is the local electric field. The voltage $V$ is obtained from the
laplacian relationship in the domain as: 

\begin{equation}
\nabla^{2}V\approx0\label{eq:Lap}
\end{equation}

where the dendrite body is part of the boundary condition per see.
The electric field is the gradient of electric potential as:

\begin{equation}
\vec{\textbf{E}}=-\nabla V\label{eq:EV}
\end{equation}
Therefore the total effective displacement $\delta\vec{\textbf{r}}$
with neglecting convection\footnote{Since Rayleigh number $Ra$ is highly dependent to the thickness (i.e.
$Ra\propto l^{3}$), for a thin layer of electrodeposition we have
$Ra<1500$ and thus the convection is negligible. \cite{Fox16}} would be:

\begin{equation}
\delta\vec{\textbf{r}}=\delta\vec{\textbf{r}}_{D}+\delta\vec{\textbf{r}}_{M}\label{eq:TotalDis}
\end{equation}
as represented in the Figure \ref{fig:Displacements}. The pulse charging
in its simplest form consists of trains of square active period $t_{ON}$
, followed by a square rest interval $t_{OFF}$ in terms of current
$I$ or voltage $V$ as shown in Figure \ref{fig:PulseCurve}. The
period $P=t_{ON}+t_{OFF}$ is the time lapse of a full cycle. Hence
the pulse frequency $f$ is:

\begin{equation}
f=\frac{1}{t_{ON}+t_{OFF}}\label{eq:Frequency}
\end{equation}
and the duty cycle\textbf{ $\mathbf{D}$} represents the fraction
of time in the period $P$ that the pulse is active :

\begin{equation}
\mathbf{D}=ft_{ON}\label{eq:DftON}
\end{equation}

While the dendrites grow, due to random nature of the evolution of
branches, they stick out randomly, which becomes a source for the
quickening growth of the dendrite per see due to the concentration
of electric field in the sharp interfaces as well as their closer
proximity to the upcoming ions. The relaxation time allows the concentrated
ions in the ramified peaks to dissipate away into the less concentrated
areas and the concentration gradient is relaxed. The time required
for the such relaxation depends on the curvature of the interface,
where the higher curvature sites would require longer time for the
concentration relaxation within the double layer region \cite{Aryanfar19Finite}.
In fact such relaxation could occur for the larger interfacial double
layer scale with the thickness of $\kappa$, spanning to the entire
cell domain, requiring the most relaxation time for the highest curvature
regions and least for the flat counterparts. From dimensional analysis
the relaxation time of the double layer in the flat electrode is in
the range of ${\displaystyle \sim\frac{\kappa^{2}}{D^{+}}}$ \cite{Hunter01}and
for the larger domain of the cell with the representative length of
$l$ would scale up to ${\displaystyle \sim\frac{l^{2}}{D^{+}}}$,
where as their geometric mean has been later considered as ${\displaystyle \sim\frac{\kappa l}{D^{+}}}$
\cite{Bazant04}. In fact the relaxation of the concentration depends
highly to the curvature of the peaks and the entire growing interface
possess a wide range of radius of curvature values $r_{d}$, expanding
from the flat surface to the highly-curved fingers as small as the
atomic value, therefore: 

\begin{equation}
r_{d}\in[r_{atom},\infty)\label{eq:rdRange}
\end{equation}

Hence, the relaxation time for a randomly-growing interface with variation
of curvature along the interfacial line, would vary as well. Hereby,
we define the feedback relaxation time $t_{REL}$ a curvature dependent
function $f(r_{d})$ multiplied by the geometric mean time for the
concentration relaxation (i.e. $RC$ time) as: 

\begin{equation}
t_{REL}=f(r_{d})\frac{\kappa l}{D^{+}}\label{eq:tREL}
\end{equation}

As the interface grows from the initial flat state ($r_{d}\to\infty$)
to creating sharp fields ($r_{d}\to r_{atom}$), the feedback relaxation
time $t_{REL}$ should adapt respectively based on the most critical
state of the interface, which is the location of ionic concentration.
Therefore the range of acceptable value for the feedback relaxation
time $t_{REL}$ should lie between the relaxation scale within the
double layer to the scale of the cell domain. The variation of the
feedback relaxation time $t_{REL}$ occurs from it's minimum value
during the instigation, to it's maximum value in the atomic scale.
Therefore considering the relevant time-scales from beginning (flat)
to the most ramified state (atomic sclae), from Equation \ref{eq:tREL}one
has: 

\begin{equation}
\begin{cases}
\lim_{r_{d}\rightarrow\infty}f(r_{d})=1 & \text{Flat}\\
\lim_{r_{d}\to r_{atom}}f(r_{d})={\displaystyle \frac{l}{\kappa}} & \text{Ramified}
\end{cases}\label{eq:BC}
\end{equation}

Assuming the form of the control function an combination of linear
and exponential terms as $f(r_{d})=ar_{d}+b\exp(cr_{d})$, from the
boundary conditions in the Equation \ref{eq:ControlFunction} one
gets: 

\begin{equation}
f(r_{d})=1+({\displaystyle \frac{l}{\kappa}}-1)\exp[-r_{d}]\label{eq:ControlFunction}
\end{equation}

Thus the feedback relaxation time $t_{REL}$ is obtained as: 

\begin{equation}
t_{REL}=\frac{\kappa l}{D^{+}}\left(1+(\frac{l}{\kappa}-1)\exp[-r_{d}]\right)\label{eq:tOpt}
\end{equation}

\begin{figure}
\noindent\begin{minipage}[t]{1\columnwidth}%
\begin{minipage}[t]{0.59\textwidth}%
\begin{center}
\includegraphics[width=1\textwidth,height=0.22\textheight]{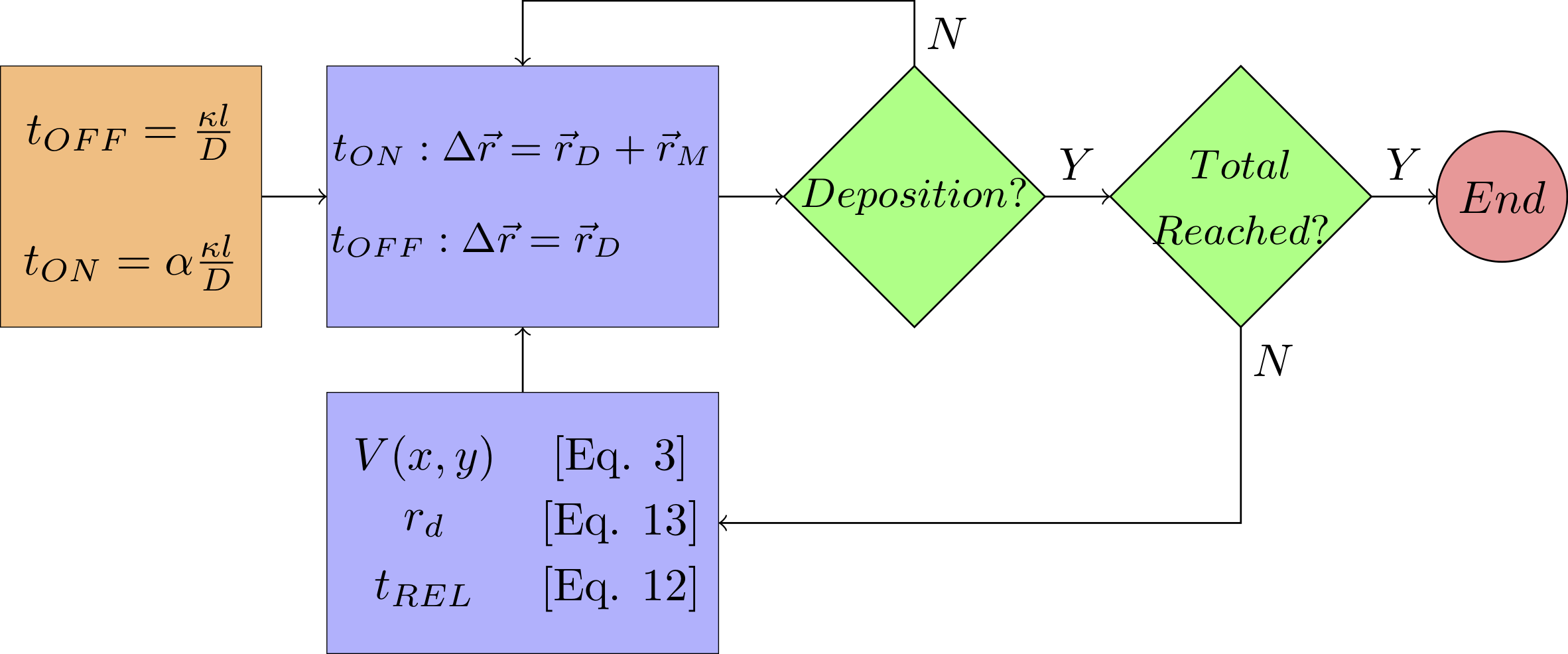}
\par\end{center}
\caption{The transport elements in the coarse scale of time.\label{fig:Flowchart}}
\end{minipage}\hfill{}%
\begin{minipage}[t]{0.4\textwidth}%
\begin{center}
\includegraphics[height=0.22\textheight]{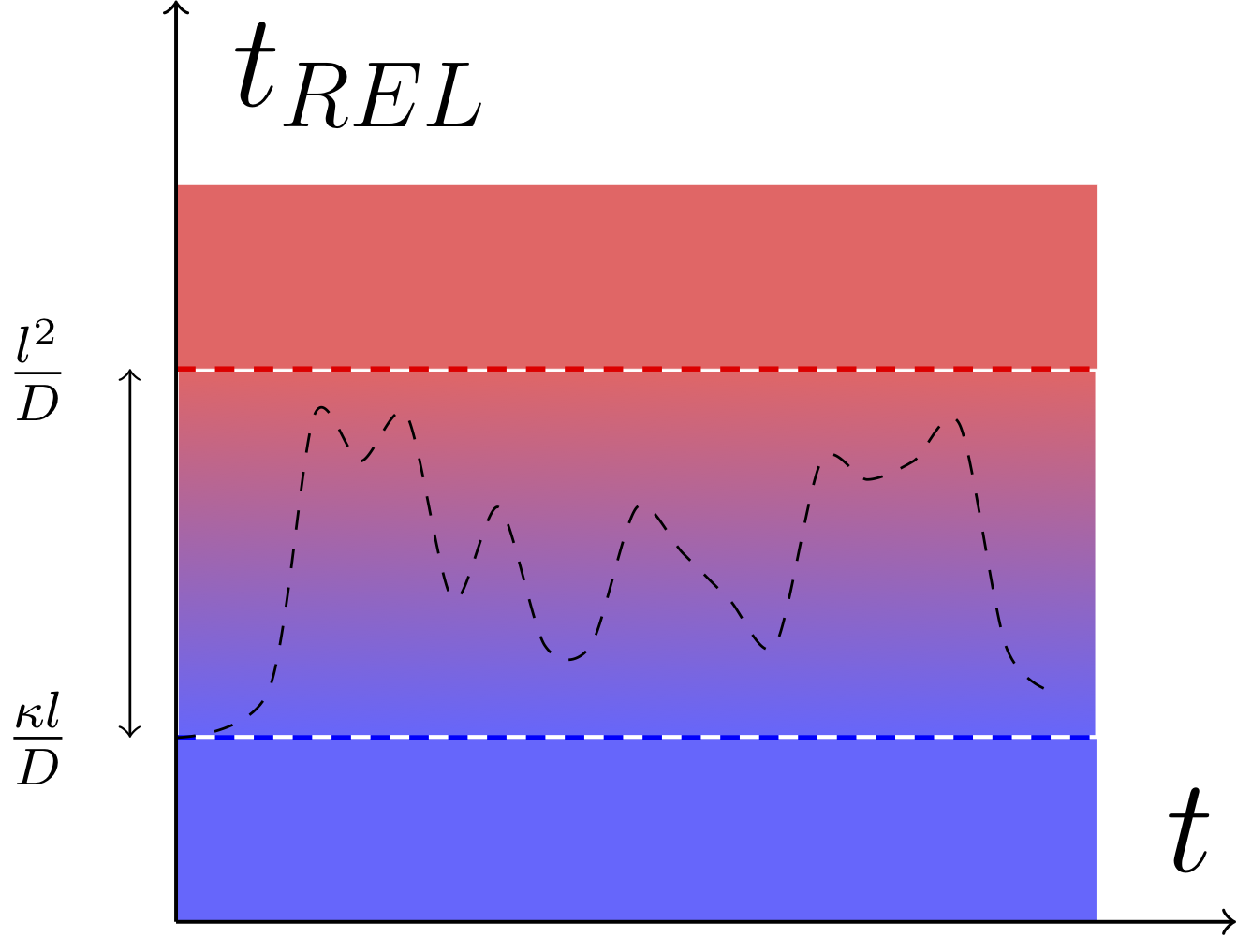}
\par\end{center}
\caption{Square pulse wave.\label{fig:tOFFControl}}
\end{minipage}%
\end{minipage}
\end{figure}

The radius of curvature $r_{d}$ can be approximated via the contours
of the iso-potential curvature of the electric field in the vicinity
of electrodeposits, where it occurs typically within the double layer
region of thickness $\kappa$. The corresponding line could be obtained
locating magnitude of the isopotential contour matching with the electrode
($V=V_{electrode}$). If $(x,y)$ represents the coordinates of the
curvature line, the point of the minimum radius of curvature would
address the most critical state, and requires higher dissipation of
ionic concentration. The radius of curvature $r_{d}$ can be calculated
from Equation \ref{eq:rd} as: 

\begin{equation}
r_{d}=min\left(\Bigg|{\displaystyle \frac{{\displaystyle 1+\frac{dy}{dx}}}{{\displaystyle \frac{d^{2}y}{dx^{2}}}}}\Bigg|\right)\Bigg|_{x=0}^{x=l}\label{eq:rd}
\end{equation}

This value is computed in real time and inserted into the feedback
algorithm. In fact the curvature-dependent relaxation time provides
a positive feedback for halting of the quickening dendrites, which
is negative feedback for dendritic evolution. The Flowchart \ref{fig:Flowchart}
represents the control loop representing the real-time computation
of the curvature and the corresponding feedback relaxation time for
the minimization of the dendritic branching. Figure \ref{fig:tOFFControl}
schematically represents such variation where the feedback relation
time $t_{REL}$ starts from the minimum value of ${\displaystyle \sim\frac{\kappa l}{D^{+}}}$
in the flat surface and varies based on the measurement of the highest
curvature of the tip given in the Equation \ref{eq:rd}. 
\begin{table}
\begin{centering}
\begin{tabular}{l|l||l||l||l|l}
Var. & \multicolumn{4}{l|}{Value} & Ref.\tabularnewline
\hline 
$\delta t(\mu s)$ & \multicolumn{4}{c|}{$1$} & \cite{Dror10}\tabularnewline
\hline 
$D(m^{2}/s)$ & \multicolumn{4}{c|}{$1.4\times10^{-14}$} & \cite{Aryanfar14Dynamics}\tabularnewline
\hline 
$\#Li^{+}$ & \multicolumn{4}{c|}{$50$} & \cite{Aryanfar14Dynamics}\tabularnewline
\hline 
$\#Li^{0}$ & \multicolumn{4}{c|}{$400$} & \cite{Aryanfar14Dynamics}\tabularnewline
\hline 
$l(nm)$ & \multicolumn{4}{c|}{$167$} & \cite{Aryanfar14Dynamics}\tabularnewline
\hline 
$\Delta V(mV)$ & \multicolumn{4}{c|}{$85$} & \cite{Aryanfar14Dynamics}\tabularnewline
\hline 
\end{tabular}
\par\end{centering}
\caption{Simulation parameters. \label{tab:Sample}}
\end{table}
The thickness of the double layer $\kappa$can be obtained from \cite{Bazant04}:

\[
\kappa=\sqrt{\frac{\varepsilon k_{B}T}{2z^{2}e^{2}C_{b}}}
\]

where $\varepsilon$ is the permittivity of the solvent, $k_{B}$
is Boltzmann constant, $T$ is the temperature, $z$ is the valence
number, $e$ is the electron charge and $C_{b}$ is the average ambient
electrolyte concentration. 

The computation was carried out based on the simulation parameters
given in the Table \ref{tab:Sample}. Figure \ref{fig:Densities}
illustrates the resulted morphologies of the grown dendrites based
on the applied relaxation time. 

The density of the electro-deposits can easily be calculated from
confining the atoms in a rectangle, the height of which spans to the
highest dendrite coordinates $h_{max}$. Therefore: 

\[
\rho=\frac{n\pi r^{2}}{h_{max}l}
\]

where $n$ is the number of atoms composing the dendrite, $r$ is
the atomic radius and $l$ is the scale of the domain. The density
of the morphologies, sample of which is shown in the Figure \ref{fig:Opt}
is provided in the Figures \ref{fig:200}, \ref{fig:400} and \ref{fig:800}
versus various relaxation time values as the control parameter. 

As well, the variation of the highest interfacial curvature (minimum
radius of curvature $r_{d}$) and their corresponding feedback relaxation
time $t_{REL}$ and the density $\rho$ versus the number of deposited
atoms is shown in the Figures \ref{fig:rd}, \ref{fig:tOFF} and \ref{fig:Density}. 

\section{Results \& Discussion}
\begin{center}
\begin{figure}
\begin{centering}
\subfloat[$t_{OFF}=0$]{\includegraphics[height=0.18\textheight]{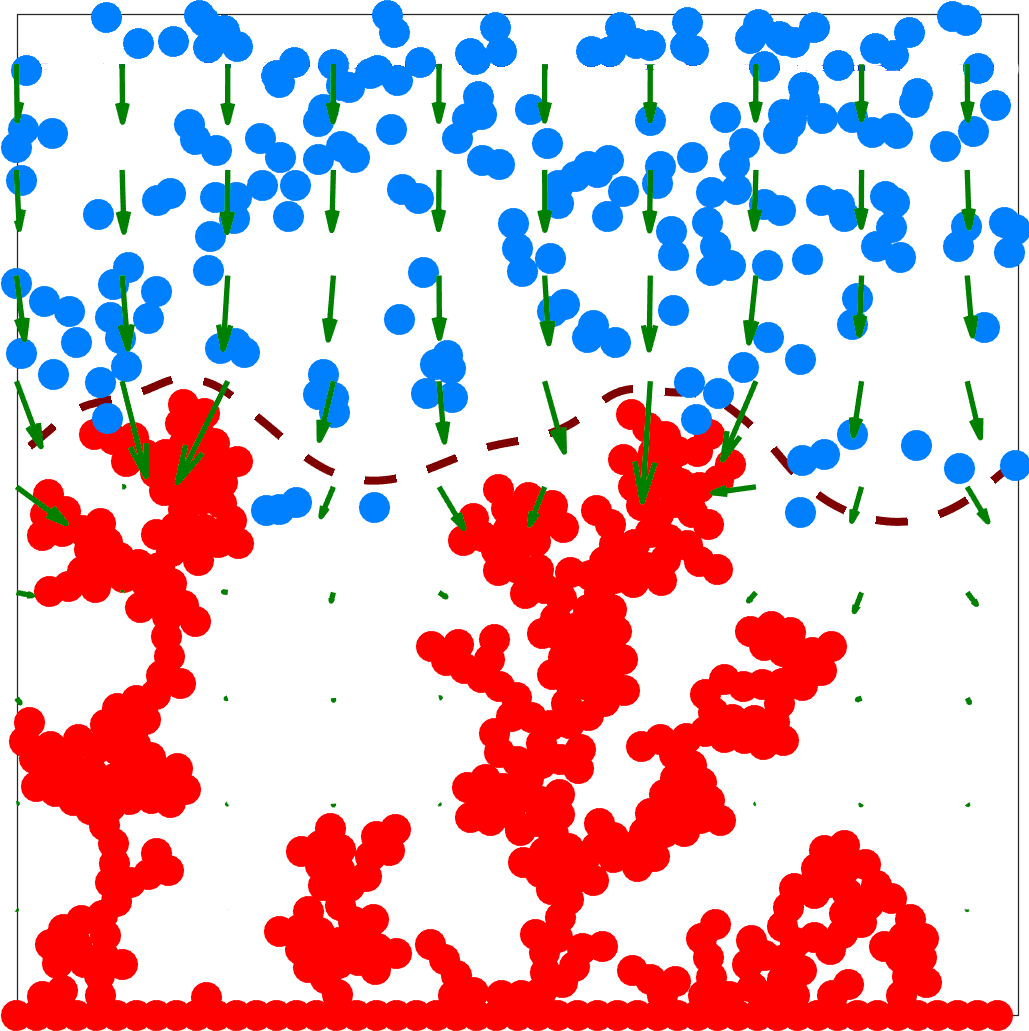}}\hfill{}\subfloat[$t_{OFF}={\displaystyle \frac{\kappa l}{D}}$]{\centering{}\includegraphics[height=0.18\textheight]{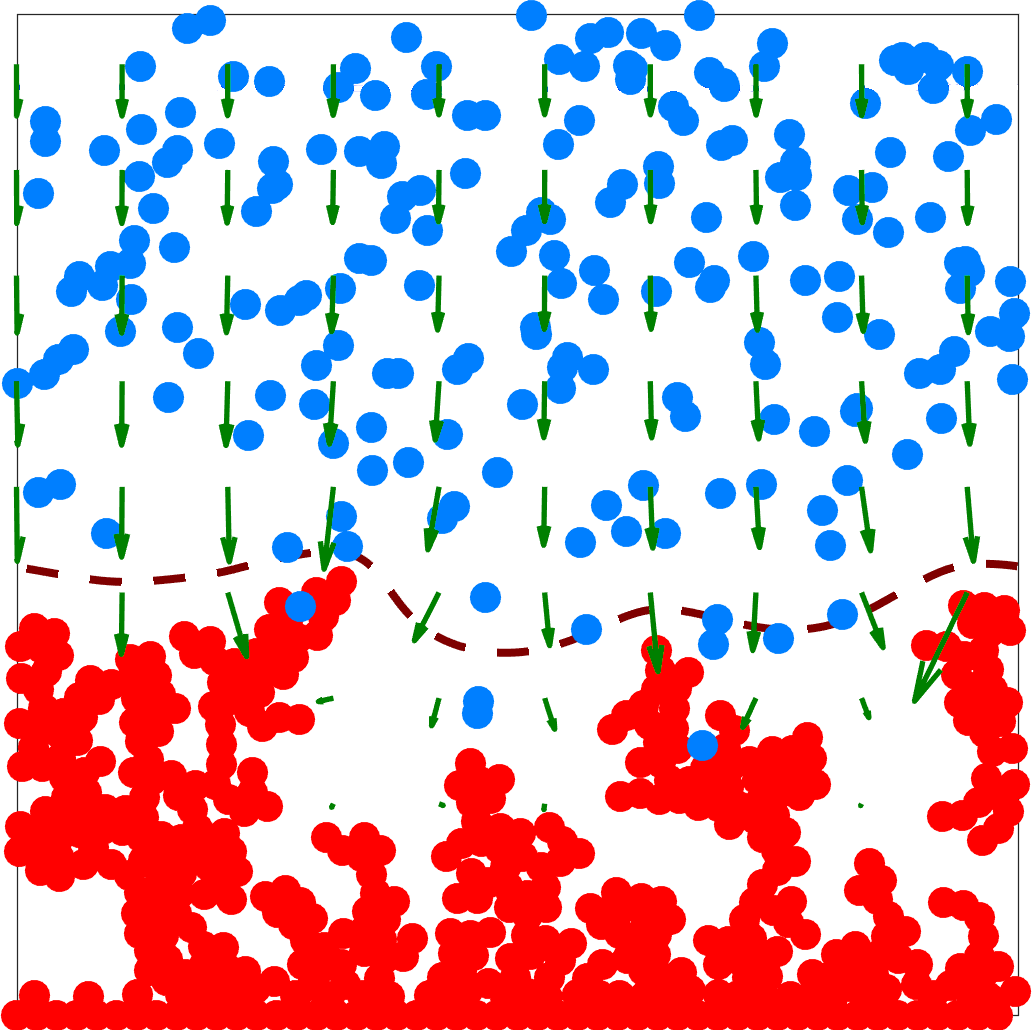}}\hfill{}\subfloat[$t_{OFF}={\displaystyle t_{Opt}}$]{\centering{}\includegraphics[height=0.18\textheight]{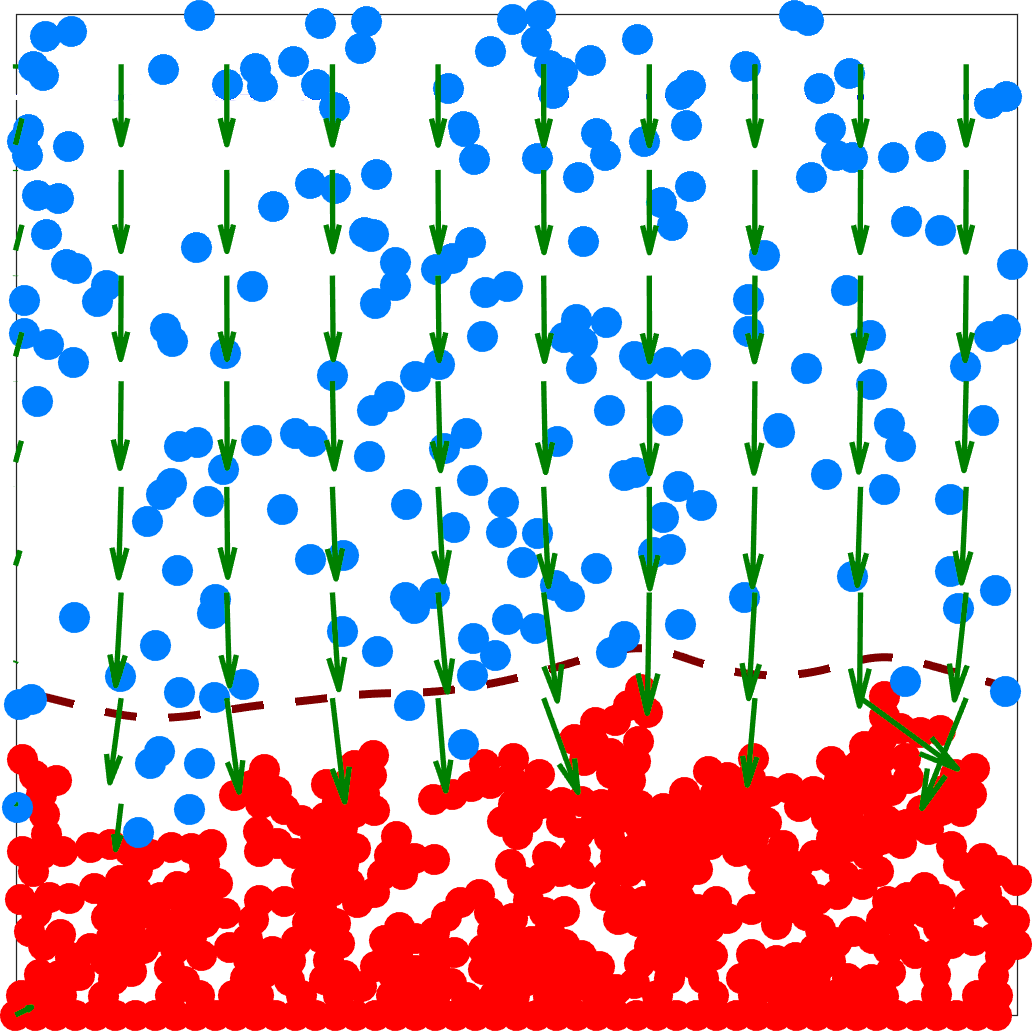}}\hfill{}\subfloat[$t_{OFF}={\displaystyle \frac{l^{2}}{D}}$]{\raggedleft{}\includegraphics[height=0.18\textheight]{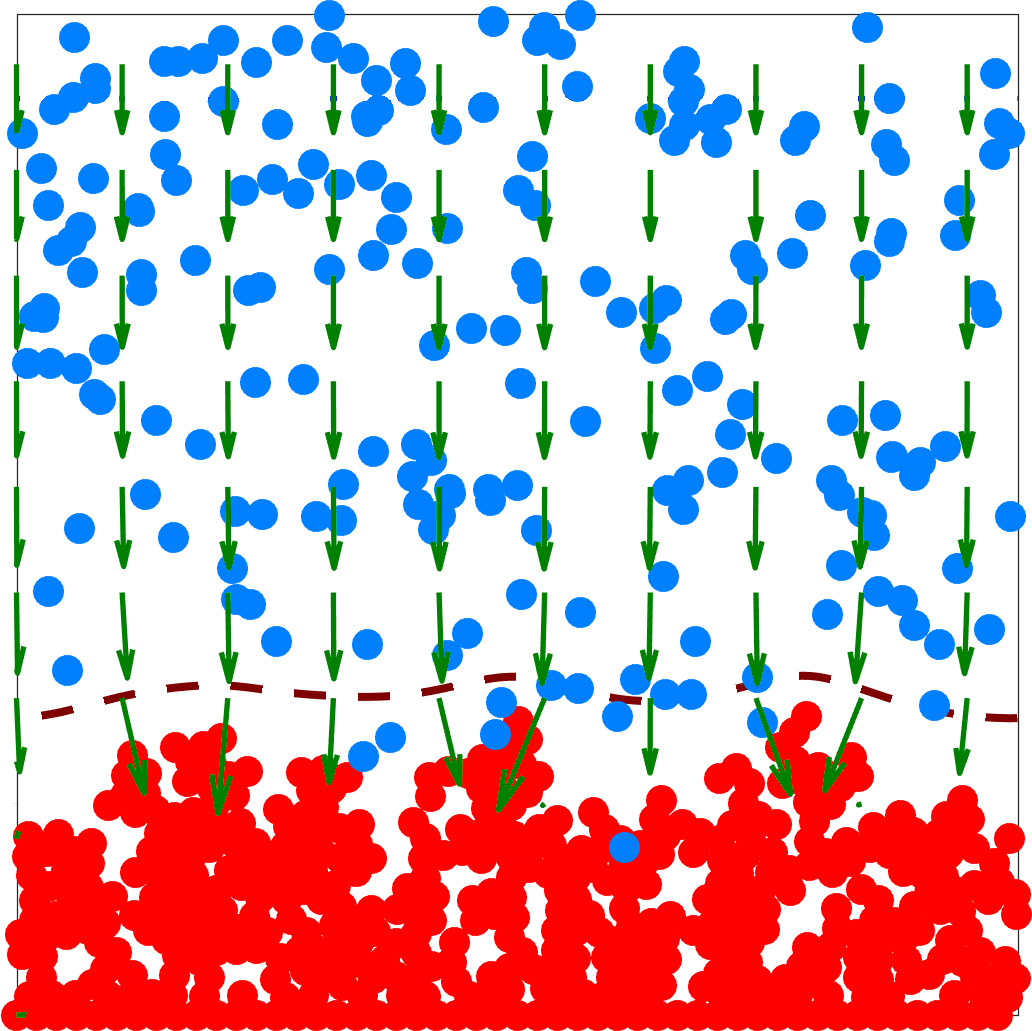}}
\par\end{centering}
\caption{Morphology variation versus the relaxation period.\label{fig:Opt}}
\end{figure}
\par\end{center}

The mechanism used in the pulse charging works based on the relaxation
of the ionic concentration in the dendritic tips. The formation of
such gradient in the concentration is, in fact, the feedback for the
quickening upcoming growth regime after the instigation and therefore
the applied feedback relaxation time should effectively dissipate
away the accumulated ions from the accumulated regions. In fact the
sharper interfaces, which have been growing faster than the rest of
the interface, have higher number of concentrated ions around them
and therefore they are in the most critical state, which have been
used for the computation of the feedback relaxation time. 

In the larger scale since the electro-migration displacement (Eq.
\ref{eq:MigDis}) scales with $\sim t$ and the diffusion displacement
(Eq. \ref{eq:DiffDis}) scales with the square of time $\sim\sqrt{t}$
. During the pulse both electro-migration and diffusion are in action
whereas during the rest period only diffusion is the main drive. Therefore
since the average reach for electro-migration is higher than the sole-diffusion
the range of reach in the rest period should in fact be competitive
with the pulse period. Therefore: 

\begin{equation}
\sqrt{2D^{+}t_{OFF}}\geq\mu^{+}\vec{\textbf{E}}t_{ON}\pm\sqrt{2D^{+}t_{ON}}\label{eq:toffton}
\end{equation}

and performing further, we get the maximum value of duty cycle $\mathbf{D}$
for effective pulse charging: 

\begin{equation}
\mathbf{D_{max}}=max\left(\frac{1}{\left(1+{\displaystyle \frac{|\vec{\textbf{E}}|}{RT}}\sqrt{{\displaystyle \frac{D^{+}}{2f}}}\right)^{2}\pm1}\right)\leq\frac{1}{2}\label{eq:Duty}
\end{equation}

where the duty cycle of the ${\displaystyle \frac{1}{2}}$ is the
limiting value for the effective suppression of dendrites. The formation
of local branches indicates that the concentration of ions in those
specific sights is high and therefore those sites should be focus
locations for the feedback relaxation time $t_{REL}$, which highly
depends on the radius of curvature $r_{d}$ of the dendritic peaks
\cite{Aryanfar19Finite}. For an individual ramified peak with the
radius of curvature $r_{d}$, the time required for the concentration
relaxation $t_{REL}^{DL}$within the double layer with the scale of
$\sim\kappa$ is : 

\begin{equation}
t_{REL}^{DL}\approx\frac{\kappa(\kappa+r_{d})}{D^{+}}\label{eq:tOFFInd}
\end{equation}

where $D^{+}$ is the diffusivity value for the ionic transport. 
\begin{center}
\begin{figure}
\subfloat[$N=200$\label{fig:200}]{\includegraphics[height=0.19\textheight]{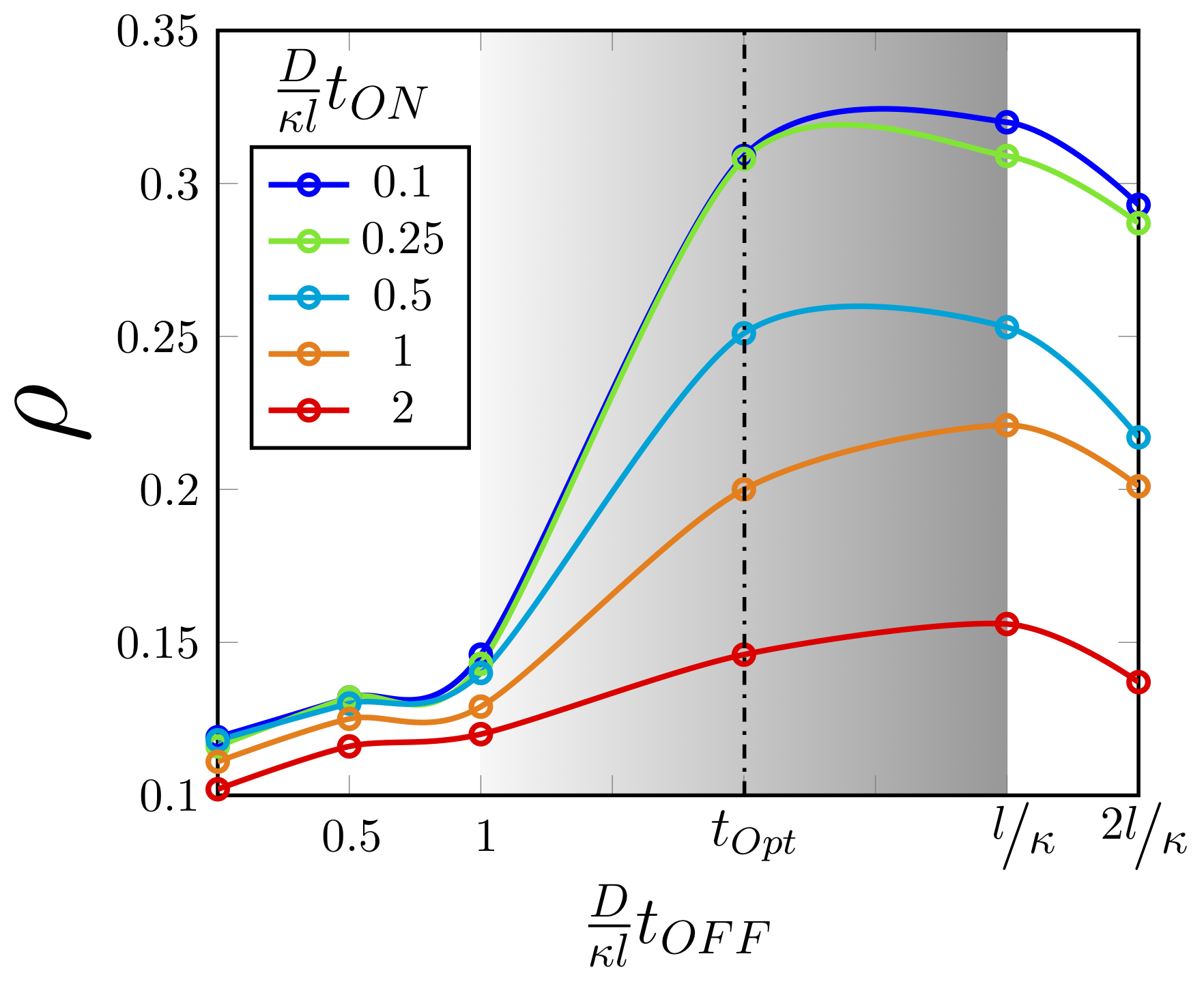}}\hfill{}\subfloat[$N=400$\label{fig:400}]{\centering{}\includegraphics[height=0.19\textheight]{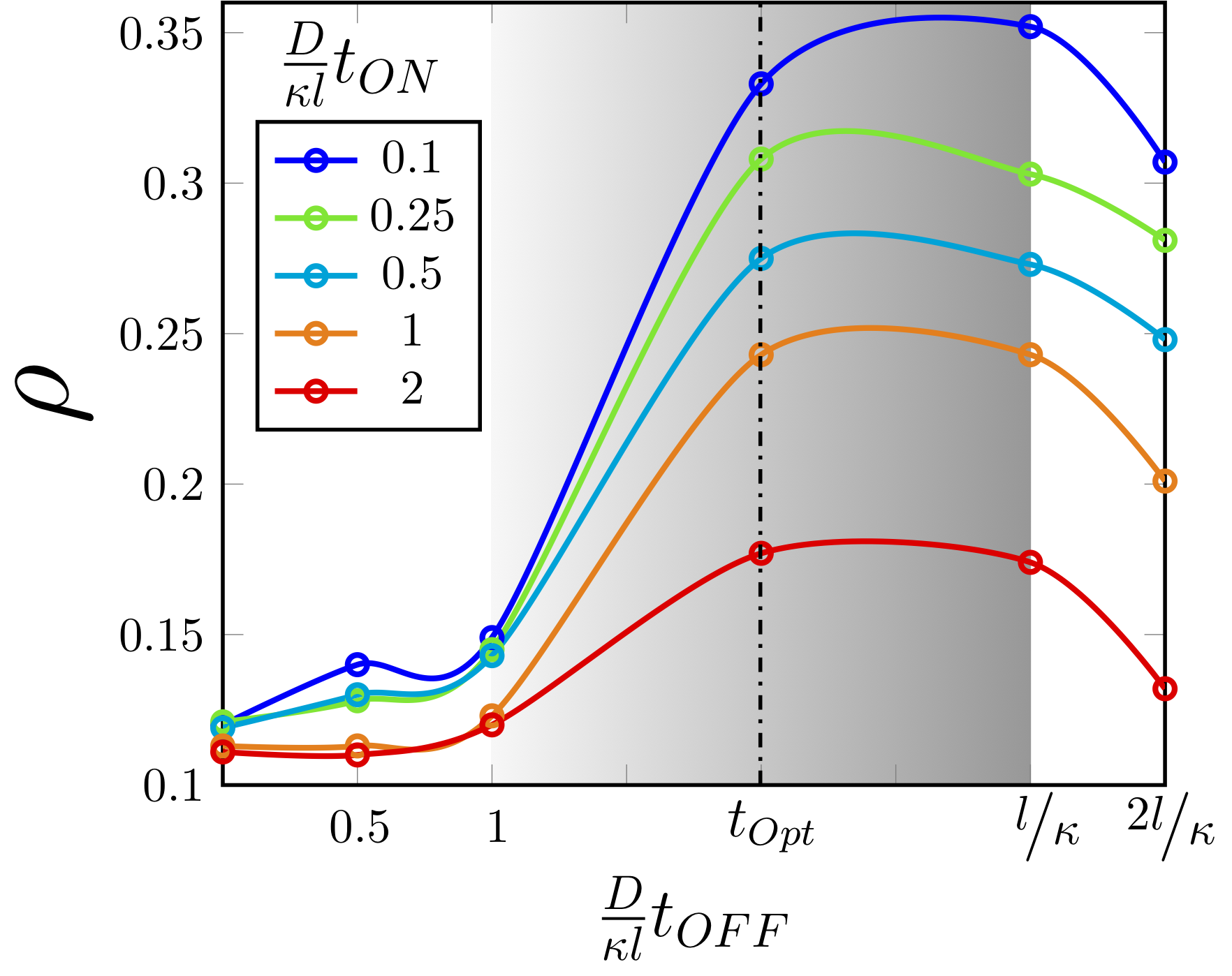}}\hfill{}\subfloat[$N=800$\label{fig:800}]{\raggedleft{}\includegraphics[height=0.19\textheight]{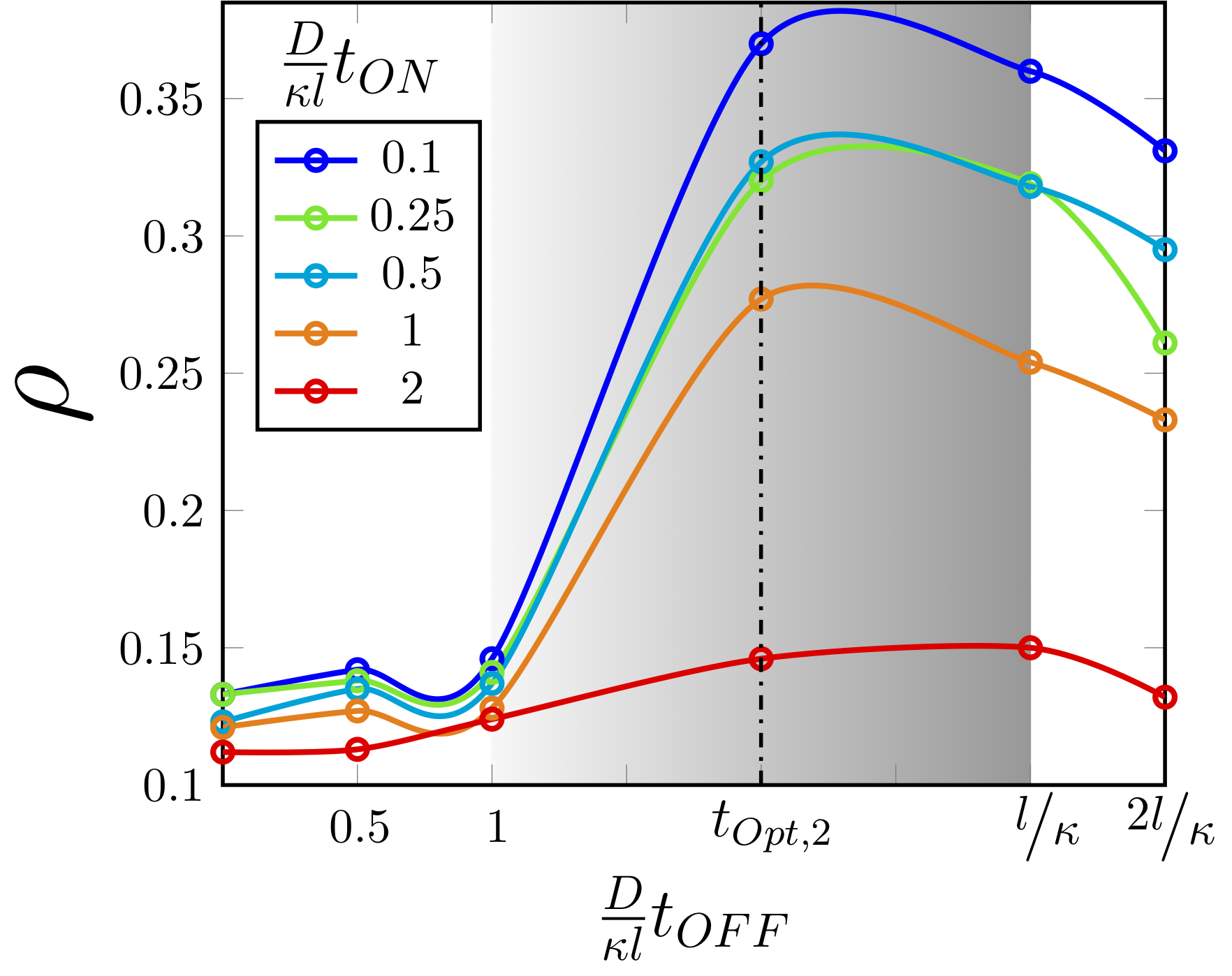}}

\caption{Density variations versus mount of atoms. \label{fig:Densities}}
\end{figure}
\par\end{center}

which in fact shows faster relaxation relative to the flat interface.
For the larger scale, extending to the entire cell domain, the feedback
relaxation time $t_{REL}$ act on the uniformization of ionic concentration
in the global range (i.e. $\sim l$) . Therefore the scale of transport
for the time and space would lead to the following comparison: : 

\begin{equation}
\frac{\kappa^{2}}{D}\leq t_{REL}^{DL}\leq\frac{kl}{D}\leq t_{REL}\leq\frac{l^{2}}{D}\label{eq:tOFFRange}
\end{equation}

Therefore the relaxation time scale varies from ${\displaystyle \sim\frac{\kappa^{2}}{D}}$
in the individual peaks to the ${\displaystyle \sim\frac{l^{2}}{D}}$
in the larger domain of the cell, and the control relaxation time
in fact varies in such range. Figure \ref{fig:Densities} represents
the density of the dendrites $\rho$ versus the intervals of pulse
$t_{ON}$ and the rest (i.e. relaxation) $t_{OFF}$ periods. The density
of the electro-deposits $\rho$correlates inversely with the pulse
charing time $t_{ON}$. This is due to the exacerbated branching during
the charge time which upon growing further gets more difficult to
halt. On the other hand, applying finer pulse periods $t_{ON}$ provides
better possibility for the suppression of dendrites. Needless to mention
that such pulse period $t_{ON}$ could not indefinitely get short
since the ions ultimately would require enough time to reach the dendrites
during this time and react from ionic to atomic species.

As well, the density values $\rho$ correlates with the relaxation
time $t_{OFF}$ until reaching a certain saturation limit. Since the
length of the domain is much larger than the double layer ($l\gg\kappa$),
the range of feedback relaxation time $t_{REL}$, shown by color gradient,
would extremely reduce the charging time with negligible compensation
in the density of electro-deposits $\rho$. The underlying reason
is that the relaxation would let the ionic concentration to relax
and uniform ionic distribution. On the other hand, extra relaxation
period will not helpful since the ionic concentration is already relaxed
and the concentration gradient has already vanished. 

As well, imposing higher-than-limit relaxation time would slightly
reduce the density $\rho$ since additional concentration from the
ambient electrolyte could be depleted in the into the non-reacting
dendritic sites. The negligible increase in the density of the dendrites
$\rho$ in the span of Figures \ref{fig:200}, \ref{fig:400} and
\ref{fig:800} illustrates the effective-ness of the pulse charging
method for the multitudes of the charge amount $N$.

\begin{figure}
\noindent\begin{minipage}[t]{1\columnwidth}%
\begin{minipage}[t]{0.32\textwidth}%
\begin{center}
\includegraphics[height=0.2\textheight]{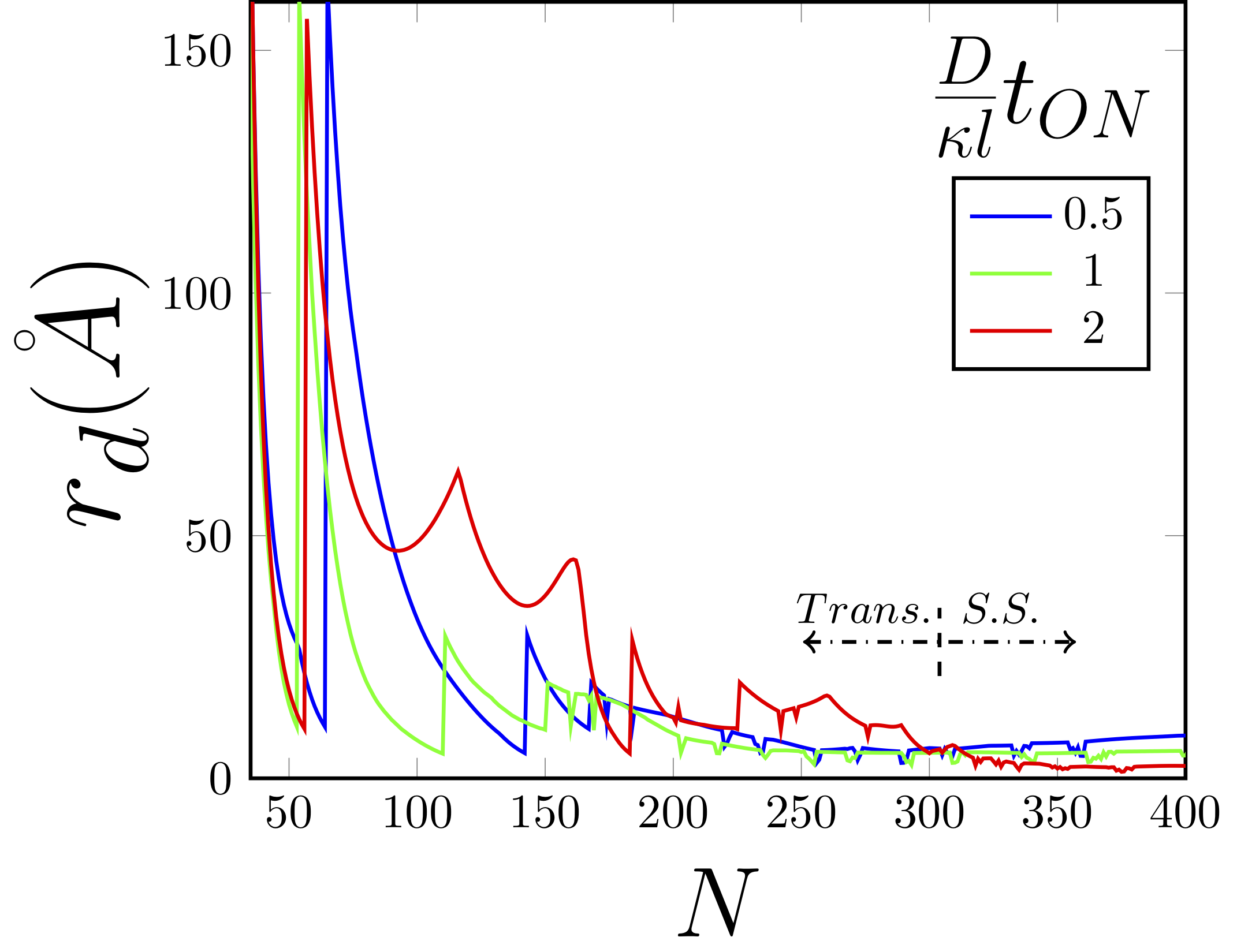}
\par\end{center}
\caption{Radius of curvature. \label{fig:rd}}
\end{minipage}\hfill{}%
\begin{minipage}[t]{0.32\textwidth}%
\begin{center}
\includegraphics[height=0.2\textheight]{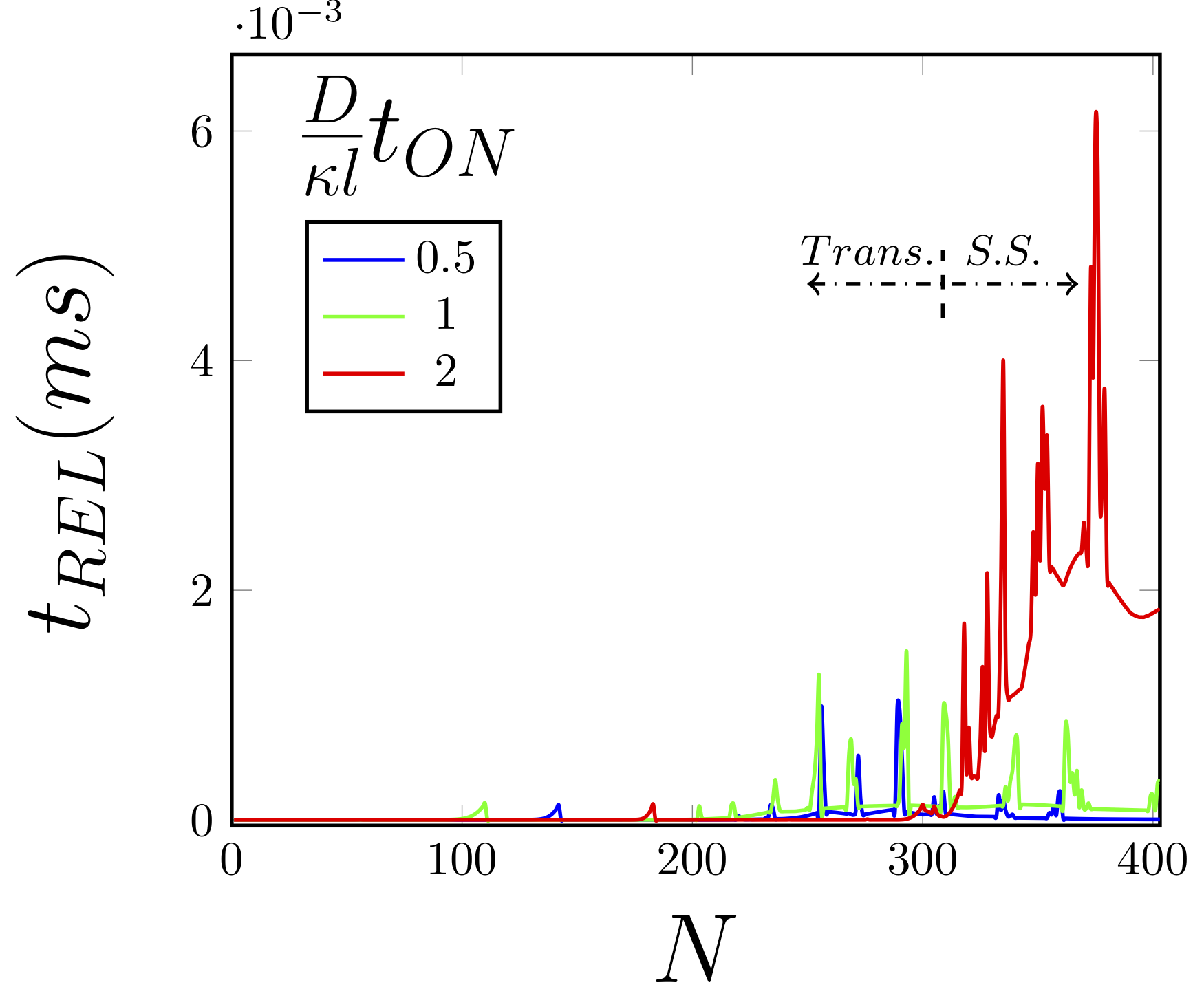}
\par\end{center}
\caption{Relaxation time. \label{fig:tOFF}}
\end{minipage}\hfill{}%
\begin{minipage}[t]{0.32\textwidth}%
\begin{center}
\includegraphics[height=0.2\textheight]{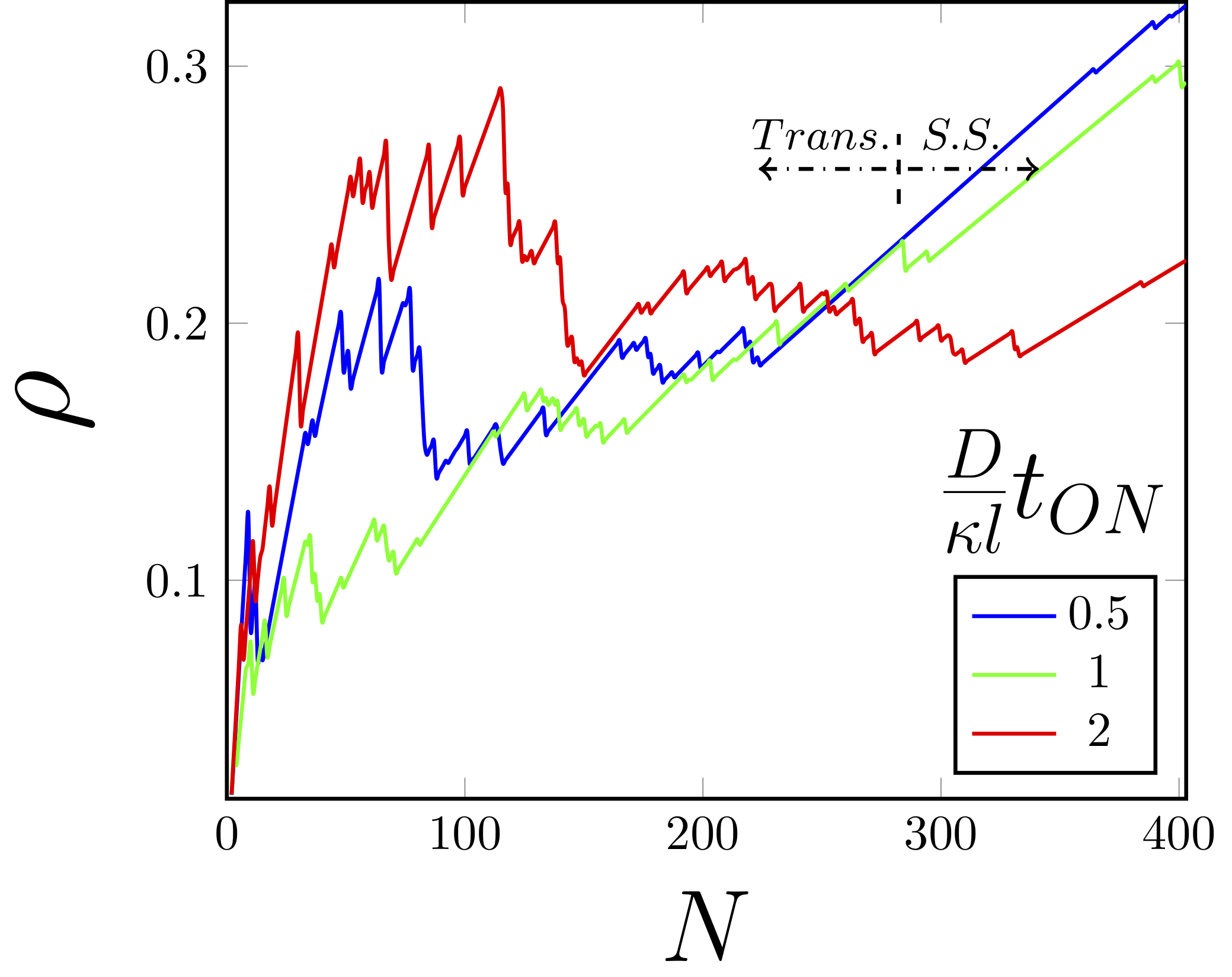}
\par\end{center}
\caption{Density. \label{fig:Density}}
\end{minipage}%
\end{minipage}
\end{figure}

The dendritic evolution can be divided into two distinctive stages
of the transient and steady-state ($S.S.$) growth regimes \cite{Bai16,Gonzalez08},
which has been illustrated in the Figures \ref{fig:rd}, \ref{fig:tOFF}
and \ref{fig:Density}. The initial transient regime in fact is a
stochastic in nature whereas the steady state regime can illustrate
an effective trend. Figure \ref{fig:rd} represents the variation
of the radius of the curvature $r_{d}$ in the growing interface versus
the deposition progress $N$ (i.e. number of the atoms). In this figure,
the transition stage during the higher pulse time shows more fluctuation
which indicates the non-uniform regime of growth for the augmented
pulse intervals. On the other hand during the steady-state ($S.S.$)
regime, radius of curvature correlates inversely with the pulse time
interval $t_{ON}$, which indicates that the morphology is controlled
for more finer pulse periods. 

Figure \ref{fig:tOFF} represents the control relaxation time $t_{REL}$
versus the progress in electrodeposition during the dendritic evolution,
where the higher pulse intervals would require higher amount of control
relaxation time $t_{REL}$ for the effective suppression of the dendrites.
As well the higher fluctuation for the higher amount of pulse charge
$t_{ON}$ shows the higher control rate due to faster dynamics of
variation in the curvature. 

The same trend of transition-to-steady state regimes has been observed
in the Figure \ref{fig:Density}, where the highest fluctuation occurs
for the higher pulsing time $t_{ON}$ where is leads to the lowest
density $\rho$ after reaching the steady growth regime. 

In fact, the quickening growth regime of the dendrites illustrates
that the larger height $h$ of the electrodeposits the rate of their
growth would be higher as well. This can simply be represented by
the following:

\[
\frac{dh}{dt}\propto h
\]

where the integration leads to the exponential relationship for the
growth regime as:

\[
h(t)\propto\exp(bt)
\]
where $b$ is the coefficient of the proportion. Setting exponential
relationship causes a very high sensitivity for the control relaxation
time $t_{REL}$ to act vigilantly versus smallest perturbation in
the ramified peaks and the form of the control relation time $t_{REL}$
proportionally contains the exponential form in Equation \ref{eq:tOpt}.
Note that other forms of the relaxation time would as well could satisfy
the boundary conditions given in the Equation \ref{eq:BC} such as
$t_{REL}^{alt}$ given as:

\begin{equation}
t_{REL}^{alt}\approx\frac{\kappa l}{D}\left(\frac{r_{d}+l}{r_{d}+\kappa}\right)\label{eq:AlternativetOFF}
\end{equation}

which has lower sensitivity for the radius of curvature $r_{d}$ relative
to proposed control relation time $t_{REL}$. This can be proven by
calculating their derivative with respect to radius of curvature  (
${\displaystyle \frac{dt_{OFF}}{dr_{d}}}$ ) and show that: 

\[
\frac{dt_{REL}}{dr_{d}}\gg\frac{dt_{REL}^{alt}}{dr_{d}}
\]

Thus from equations \ref{eq:tOpt} and \ref{eq:AlternativetOFF} and
considering the negative value for both derivatives, one must have: 

\begin{equation}
{\displaystyle \frac{\kappa-l}{\kappa\exp(r_{d})}}\frac{\kappa l}{D}>\frac{\kappa-l}{(r_{d}+\kappa)^{2}}\frac{\kappa l}{D}\label{eq:Comparison}
\end{equation}

since $\kappa\ll l$ dividing by negative value of $\kappa-l$ changes
the inequality sign, therefore: 

\begin{equation}
\exp(r_{d})>(r_{d}+\kappa)^{2}\label{eq:ExpQuad}
\end{equation}
The equation \ref{eq:ExpQuad} is obvious for a large values of radius
of curvature $r_{d}$ since the exponential term in the denominator
will surpass the quadratic term in the right side. As well for an
infinitesimally small value of the radius of curvature $r_{d}$ one
can use Taylor expansion as: $\exp(r_{d})\approx1+r_{d}{\displaystyle +\cancelto{0}{O(r_{d}^{2})}}$
and one has: 

\[
\kappa(1+r_{d})>(r_{d}+\kappa)^{2}
\]

re-arranging gives: 

\[
r_{d}^{2}+\kappa r_{d}+\kappa^{2}-\kappa<0
\]

which is a quadratic equation in terms of the radius of curvature
$r_{d}$ and the root is found as: 

\[
r_{d}=\sqrt{\kappa-\frac{3}{4}\kappa^{2}}-\frac{\kappa}{2}
\]

Considering the infinitesimal value for thickness of the double layer
($\kappa\to0$) the value for $r_{d}$ would be very small. Therefore
for the most range of $r_{d}\in[r_{atom},\infty)$ the exponential
relationship remains as the most sensitive to the variations in the
radius of curvature $r_{d}$ and the Equation \ref{eq:tOpt} as effective
control relaxation time for suppression of the dendrites.. 

In practice, the discerning the formation of such a peak in the dendrite
morphology for potentiostatic charging (constant applied voltage $V$)
could be obtained by computing the sudden increase in the current
density, representing the runaway process, whereas for galvanic charging
(constant applied current $I$) could be the sudden drop in the potential
value, where both of these events represent the runaway process (i.e
jump) in time.. 

\section{Conclusions}

In this paper we have developed an effective real-time feedback control
relaxation method for minimization of dendritic grown during electrodeposition
for preventing the branched evolution of the grown microstructures.
The control parameter has been considered as the maximum curvature
of the growing interface based on the radius of curvature of the most
critical peaks. The sensitivity of the feedback relaxation time to
the curvature has been analyzed to be extremely high with the exponential
correlation, analogous to the growth dynamics of the branches. The
methodology can be used for smart charging in rechargeable batteries
for the controlling the morphology of the grown electro-deposits. 

\bibliographystyle{unsrt}
\bibliography{Refs}

\end{document}